\documentstyle[12pt]{article}
\date{June 13, 1997}

\title{Polarization of high-energy electrons
traversing a laser beam}

\author{G.L.~Kotkin$^{a)}$\thanks{Electronic address:
kotkin@phys.nsu.nsk.ru},
$\;$H.~Perlt$^{b)}$\thanks{Electronic address:
perlt@tph204.physik.uni-leipzig.de}
$\;$ and V.G.~Serbo$^{a)}$\thanks{Electronic address: serbo@math.nsc.ru}\\
{\it $^{a)}$Novosibirsk State University, 630090, Novosibirsk, Russia}\\
{\it $^{b)}$Institut f\"ur Theoretische Physik, Leipzig University,
04109, Leipzig, BRD}}

\begin{document}

\maketitle

\begin{abstract}
When polarized electrons traverse a region where the laser
light is focused their  polarization  varies 
even if their energy and direction of motion are not changed.
This effect is due to interference of the incoming electron
wave and an electron wave scattered at zero angle. Equations are obtained
which determine the variation of the electron density matrix, and
their solutions are given. The change in the electron
polarization depends not only on the Compton cross section but
on the real part of the forward Compton amplitude as well. It should be 
taken into
account, for example, in  simulations of the $e \to \gamma$
conversion for future $\gamma \gamma$ colliders.
\end{abstract}

\section{Introduction}

Let an electron beam traverse a region where the laser light is
focused. It is well known that the energies of these electrons as well as their
polarizations are varied due to Compton scattering. The
polarization of the scattered electrons in dependence on the spin
states of the colliding electrons and photons was discussed in
Refs. \cite{Tol,Groz,KPS}.

However, when the electron passes the laser beam the polarization varies also
for those electrons which conserve their energies and
directions of motion. In the present paper we calculate the
polarization of  those undeflected electrons. To our knowledge 
this variation of polarization has been considered 
in previous papers on this subject  as a 
consequence of the fact that electrons with different values of
their helicities are knocked out of the beam differently
\footnote{For example, this effect was considered as the base for
obtaining the polarized electron beams in Ref. \cite{Derb}.}.
In the present paper we use a more general approach. We consider
the variation in polarization of the undeflected electrons as the
result of interference of the incoming electron wave and the wave
scattered at zero angle. This approach allows to reproduce the previous
results, but we found that they are not
complete. The reason is that previous authors  consider the 
electron state as a
mixture of the incoherent helicity states. However, for electrons
possessing a transverse polarization there is a defined phase relation between the
helicity states. Moreover, the
coherence of helicity states is not destroyed in the scattering
process which leads to the rotation of the transverse component
of the electron polarization vector as well as to variation in
its absolute value. This phenomenon is similar to the well-known effect of
the rotation of the neutron spin in the polarized target
\cite{Barysh}.

The topic of this paper is to be seen in connection with projects of
$\gamma \gamma$ colliders which are under development now (see Refs.
\cite{gg,Berkeley,ZDP,TESLA}). In these projects it is suggested to
obtain the required high-energy $\gamma$-quanta by backward
Compton scattering of laser light on the electron beam of a
linear collider. The spectrum of the obtained $\gamma$-quanta
depends very strongly on the electron polarization, therefore, it
is necessary to take into account the variation in polarization
of the undeflected electrons in  simulations of the $e \to \gamma $
conversion as well as of the $\gamma \gamma$ luminosity.

In the next section we derive the equations for components of
the electron polarization vector. They are suited to
simulation of the $e \to \gamma$ conversion at the $\gamma
\gamma$ colliders. These equations include not only the Compton
cross sections (which are proportional to the imaginary parts of
the corresponding forward amplitudes) but the real parts of the
same amplitudes as well. In the last section we give solutions
to these equations, compare with  previous results and
discuss a possible scheme of simulation.

Note, that with the growth of intensity of laser flash it is
necessary to take into account the effects of the intense
electromagnetic fields (see Refs. \cite{NR,BKMS} and literature
therein) which we neglect in present paper.

\section{Equations for the electron polarization vector }

In this section we derive the equations for the electron polarization
vector based on the  method  given in \cite{KS} where it was used to solve the 
analogous
problem of polarization of 
$\gamma$-quanta traversing a laser
bunch.

Let us consider the head-on collision of electrons with the bunch
of laser photons. We choose the $z$ axis along the momenta of
electrons. The polarization state of an electron is described by
the polarization vector $\mbox{\boldmath $\zeta$}$. The electron
density matrix in the helicity basis ($\lambda,\; \lambda' = \pm
{1\over 2}$) has the form
\begin{eqnarray}
\rho _{\lambda \lambda '}^e = {1\over 2}
\left(
\begin{array}{cc}
1+\zeta_z           &  \zeta_x -i \zeta_y \\
\zeta_x + i \zeta_y  &  1-\zeta_z
\end{array}
\right) \,.
\label{2}
\end{eqnarray}

For a laser photon the density matrix is described by the following
parameters: the degree of the circular polarization $P_c$, the
degree of the linear polarization $P_l$ and the angle $\gamma$ of
the linear polarization direction. In the helicity basis
($\lambda,\; \lambda' = \pm 1$) this matrix has the form (see,
for example, Ref. \cite{BLP} \S 8):
\begin{eqnarray}
\rho _{\lambda \lambda '}^L = {1\over 2}
\left(
\begin{array}{cc}
1+P_c & -P_l \,{\rm e}^{-2i\gamma} \\
-P_l \,{\rm e}^{2i\gamma}  & 1-P_c
\end{array}
\right) \,.
\label{3}
\end{eqnarray}

We will use also  a compact expression describing the polarization
of the electron and  photon
\begin{equation}
\rho_{\Lambda \Lambda'} = \rho^e_{\lambda_1\lambda_1'}\;
\rho^L_{\lambda_2 \lambda_2'}\,.
\label{4}
\end{equation}

In the following we derive equations for components of the
polarization vector of an electron traversing a laser bunch. As
is well known the variations in intensity and polarization of the
wave passing through a medium are due to interference the
incoming wave with the wave scattered at zero angle. Let the
incoming wave have the form
$$
A_{\Lambda}\,{\rm e}^{ikz}\,.
$$
Here the amplitude $A_{\Lambda}$ describes the polarization state
of the electron with the energy $E$ and the laser photon with the
energy $\hbar \omega \ll E$, the wave vector being
$k=\sqrt{E^2-m^2c^4} /(\hbar c)$.  When the wave passes through a
``target" layer of a thickness $dz$ and density $n_L$ a forward scattered wave
appears
\begin{equation}
f_{{\Lambda \Lambda'}} \,A_{\Lambda'}\,2n_L \, dz\, \int { {\rm
e}^{ikr} \over r}\, dx\, dy =
{2\pi \, i \over k}\, f_{\Lambda \Lambda'} \,A_{\Lambda'}\,2n_L
\, dz\, {\rm e}^{ikz} \, =
\,{\rm e}^{ikz}\, dA_{\Lambda} \,,
\label{5}
\end{equation}
where $f_{{\Lambda \Lambda'}}$ is the forward amplitude for the
process of the Compton scattering. The factor 2 in front of $n_L$
is due to relative motion of the electrons and the ``target".

The matrix $\rho_{{\Lambda \Lambda'}}$ from Eq. (\ref{4}) is
expressed via the product of $A_{\Lambda}$:
\begin{equation}
\rho_{\Lambda \Lambda'} = {\langle A_{\Lambda} A^*_{\Lambda'}
\rangle \over B } \,, \;\;\;\;\;
B=     \langle A_\Lambda A^*_{\Lambda} \rangle \, ,
\label{6}
\end{equation}
where  $\langle ... \rangle$ denotes a statistical averaging. The
quantity $B$ is proportional to the number of electrons $N_e$.
When the electron wave passes through the layer of thickness
$dz$ its relative variation in intensity is equal to
\begin{equation}
{dN_e\over N_e} ={dB\over B}= {2\over B} {\rm Re}\, \langle \,
(dA_\Lambda )\, A^*_{\Lambda} \rangle \, = -{4\pi \over k} {\rm
Im}\,(f_{\Lambda \Lambda'} \, \rho_{\Lambda' \Lambda}) \, 2n_L\,
dz \,.
\label{7}
\end{equation}
This equation can be cast into the form
\begin{equation}
dN_e = -\,\sigma_{C}\,2n_L dz \, N_e\,,
\label{8}
\end{equation}
where
\begin{equation}
\sigma_{C} = {4\pi \over k}\, {\rm Im}\, (f_{\Lambda
\Lambda'} \, \rho_{\Lambda' \Lambda}) \,
\label{9}
\end{equation}
is the total Compton cross section (optical theorem).
Analogously,
\begin{equation}
d\rho_{\Lambda \Lambda'} = d\, {\langle A_{\Lambda} A^*_{\Lambda'}
\rangle \over B } \, = {2\pi\,i \over k}
(f_{\Lambda \Lambda''}\rho_{\Lambda'' \Lambda'} -
f^*_{\Lambda' \Lambda''}\rho_{\Lambda \Lambda''}) \, 2n_L dz-
\rho_{\Lambda \Lambda'}\, {dB\over B}\,.
\label{10}
\end{equation}

The scattering amplitudes $f_{\Lambda \Lambda'}$ can be expressed by
the relativistic helicity amplitudes $M_{\lambda_1 \lambda_2\,
\lambda'_1 \lambda'_2}$
$$
f_{\Lambda \Lambda'} \equiv
f_{\lambda_1 \lambda_2\, \lambda'_1 \lambda'_2}=
{k \lambda_C^2\over 4\pi x} \,
M_{\lambda_1 \lambda_2\, \lambda'_1 \lambda'_2}\,,
$$
\begin{equation}
\lambda_C={\hbar  \over m c},\;\;\;\;
x ={s-4m^2 c^4 \over m^2 c^4} ={2 (E +\hbar k c)\, \hbar \omega_L
\over m^2 c^4}\,,
\label{11}
\end{equation}
where $m$ is the electron mass, $\lambda_C$ is the electron
Compton wavelength  and $s$ is the square of the total energy of
the electron and the laser photon in their centre-of-mass system.
Among the four independent helicity amplitudes for the Compton
process only two are not equal zero for the forward
scattering (compare \cite{BLP} \S 70):
$$
M_{++++}=M_{----}\,,\;\; M_{+-+-}=M_{-+-+}\,.
$$

According to the optical theorem, the imaginary parts of these
amplitudes are connected with the Compton cross sections
$\sigma_{++}$ and $\sigma_{+-}$ for collisions of electrons and
photons with helicities $\lambda_1 = + {1\over 2}, \; \lambda_2 =
+1$ and $\lambda_1 = + {1\over 2}, \; \lambda_2 = -1$,
respectively:
\begin{equation}
\sigma_{++}={\lambda_C^2 \over x }\,{\rm Im}\, M_{++++}\,, \;\;\;
\sigma_{+-}={\lambda_C^2\over x}\,{\rm Im}\, M_{+-+-}\,.
\label{12}
\end{equation}

Instead of these two amplitudes, it is convenient to use their
half-sum and half-difference. The imaginary part of the first
quantity is related to the Compton cross section for 
unpolarized particles
\begin{equation}
\sigma_{np} = {1\over 2} (\sigma_{++} + \sigma_{+-})  = \pi
r^2_e\, I_{np}\, ,
\label{13}
\end{equation}
where $r_e =e^2/(m c^2)$ is the classical electron radius.

Further, instead of half-difference we will use real
dimensionless quantities $R$ and $I$ proportional 
to the real and imaginary parts of half-difference of the forward
scattering amplitudes, correspondingly
\begin{equation}
\pi r_e^2\,( R +i I)={\lambda_C^2 \over x }\,{1\over
2}\,(M_{++++}\, -\, M_{+-+-}) \, .
\label{14}
\end{equation}
Note that
\begin{equation}
\pi r^2_e \, I = {1\over 2} (\sigma_{++} -\sigma_{+-})\,.
\label{15}
\end{equation}

By substituting Eqs. (\ref{4}), (\ref{11})-(\ref{14}) into Eqs.
(\ref{7}) and (\ref{10}) and after summing up over the
polarization states of the final photons, we obtain equations for the
components of the electron polarization vector. To write down
these equations it is convenient to introduce the quantity $t$ via
the relation
\begin{equation}
dt(z) = 2\pi r^2_e \, n_L dz.
\label{16}
\end{equation}
$dt$ is called the reduced optical thickness of the layer
$dz$. Then
\begin{eqnarray}
{d\zeta_x\over dt} &=& (R \zeta_y +I \zeta_z \zeta_x) \,P_c, \nonumber\\
{d\zeta_y\over dt} &= &(-R \zeta_x +I \zeta_z \zeta_y) \,P_c, \\
{d\zeta_z\over dt} &= &- I (1- \zeta_z^2)\, P_c \nonumber .
\label{17}
\end{eqnarray}
Let us note that these equations do not depend on the degree of the
linear polarization of the laser photon $P_l$. Integrating this
system of equations one obtains the dependence of the electron
polarization vector on the reduced optical thickness $t$.

With theses results one calculates the $t$-dependence of cross section (\ref{9}) 
\begin{equation}
\sigma_{C} = \pi r^2_e\,( I_{np} + \zeta_z P_c \, I)\, ,
\label{18}
\end{equation}
and then the number of electrons $N_e(t)$ from Eq. (\ref{8}) .

The forward scattering amplitudes and, therefore, the quantities
$I_{np},\; R$ and $I$ depend on the single variable $x$
(\ref{11}) only. Let us write down formulas for $I_{np}$ and $I$
(see, for example, Ref. \cite{Tol}):
$$
\sigma_{np}= \pi r^2_e\, I_{np} = {2\pi r^2_e\over x} \left[
\left( 1- {4\over x} -{8\over x^2} \right)\, \ln{(x+1)} + {1\over
2}+ {8\over x} - {1\over 2(x+1)^2} \right] \, ,
$$
\begin{eqnarray}
\pi r^2_e\, I &=& {1\over 2} (\sigma^{++} -\sigma^{+-}) \nonumber \\
              &=& {2\pi r^2_e\over x} \left[ \left( 1+ {2\over x} \right)\,
\ln{(x+1)} - {5\over 2}+ {1\over x+1} - {1\over 2(x+1)^2} \right]
\, .
\label{19}
\end{eqnarray}

The Compton forward amplitude in the Born approximation 
(${\cal O}(\alpha)$ ) is equal to
\begin{equation}
M_{\lambda_1 \lambda_2\,\lambda'_1 \lambda'_2} =
-8\pi \, \alpha \, \delta_{\lambda_1 \lambda'_1}\,
\delta_{\lambda_2 \lambda'_2}
\label{20}
\end{equation}
and, therefore, the difference $M_{++++}\, -\, M_{+-+-}$ is
equal to zero in this approximation.

In order ${\cal O}(\alpha^2)$ the quantity $R$ is given in Ref.
\cite{BKMS} as
\begin{eqnarray}
R(x)&=& {2\over \pi\, x} \left[
         \left( 1- {2\over x} \right) F( x-1)
       - \left( 1+ {2\over x} \right) F(-x-1)\right. \nonumber \\
  & & \left. -{2x^3 \ln{x} \over (x^2-1)^2}
+ {x\over x^2-1}-{2\pi^2\over 3x} \right]  ,
\label{22}
\end{eqnarray}
where
\begin{equation}
F(x) \, = \,
\int\limits_0^x \, {\ln{|1+t|}\over t} \, dt\,
\label{23}
\end{equation}
is the Spence function. We have checked that the same function
$R(x)$ can be obtained by  the following dispersion
relations:
\begin{equation}
R(x) = {2\over \pi} {\cal P} \int\limits_0^\infty \,
{x^\prime I(x^\prime)\over x^{\prime \,2} -x^2}\, dx^\prime \, ,
\label{21}
\end{equation}
where ${\cal P}$ means the principal value of the integral.

Let us give some particular values of the
function $R(x)$:
$$
R(x)=  {x^2\over \pi}\, \left( {10\over 3}
\ln{{1\over x}} - {37\over 18} \right) \;\;\;
{\rm ¯à¨} \;\; x\ll
1 \, ,\;\;\;
R(x)=  {\pi\over x } \;\;\; {\rm ¯à¨} \;\; x \gg 1 \, ,
$$
\begin{equation}
R(1) = {\pi^2 -6 \over 6 \pi} = 0.205 \, ,\;\;\;\;
R(4.8)=0.223 \,.
\label{24}
\end{equation}
The functions $R$ and $I$ are plotted in Fig. 1.

\section{Discussion}

{\bf 1}. The solution of Eqs. (\ref{17}) has the form
\footnote{We restrict ourselves to the case when the degree of
the circular polarization $P_c$ is constant inside the laser
flash.}:
\begin{eqnarray}
\zeta_x &=& {\zeta_x^0 \cos{\varphi}+\zeta_y^0
\sin{\varphi} \over \cosh{\tau} - \zeta_z^0\, \sinh{\tau}}, \;
\zeta_y = {-\zeta_x^0 \sin{\varphi}+\zeta_x^0
\cos{\varphi}\over \cosh{\tau} - \zeta_z^0\, \sinh{\tau}},\\
\zeta_z &=& {\zeta_z^0 \cosh{\tau} - \sinh{\tau} \over \cosh{\tau}
- \zeta_z^0\, \sinh{\tau}}\nonumber
\label{25}
\end{eqnarray}
where
\begin{equation}
\varphi= P_c\,R\, t\, , \;\;\;
\tau = P_c\,I\,t\,,
\label{26}
\end{equation}
and $\zeta_x^0 ,\; \zeta_y^0, \; \zeta_z^0$ are the initial values of 
the components of
the electron polarization vector. If we introduce the auxiliary
quantities $\zeta_{\bot}^0\,\; \zeta_{\bot} ,\;\varphi_0$,  ¨
$\tau_0$ defined as
\begin{eqnarray}
\zeta_x^0 &=& \zeta_{\bot}^0\,\cos{\varphi_0}, \; 
\zeta_y^0 = \zeta_{\bot}^0\,\sin{\varphi_0}, \nonumber\\
\zeta_z^0 &=& \tanh{\tau_0}, \;
\zeta_{\bot}(\tau ) = \zeta_{\bot}^0\,{\cosh{\tau_0}\over
\cosh{(\tau_0-\tau)}}
\label{27}
\end{eqnarray}
then Eqs. (\ref{25}) can be cast into a form convenient for
the analysis
\begin{equation}
\zeta_x= \zeta_{\bot}\, \cos{(\varphi_0 -\varphi)}, \;\;\;
\zeta_y=\zeta_{\bot}\,\sin{(\varphi_0 -\varphi)}, \;\;\;
\zeta_z=\zeta_z^0\,\tanh{(\tau_0-\tau)} \,.
\label{28}
\end{equation}
It is seen from these solutions that the transverse electron polarization
($\zeta_x, \zeta_y$) rotates on the angle $(-\varphi)$ with increasing
optical thickness $t$. Its magnitude varies nonmonotonically and tends
to zero at large $t$. The longitudinal component
$\zeta_z$  varies monotonically and  tends to $\pm 1$ at $\tau \to
\mp \infty$.

Let us consider now the number of the undeflected electrons
$N_e(t)$. Substituting Eqs. (\ref{25}) and (\ref{18}) into
(\ref{8}), we obtain
\begin{equation}
N_e= N_{e}^0\,({\rm ch}\,\tau - \zeta_z^0\, {\rm sh}\,\tau )\,
{\rm e}^{-I_{np}t},
\label{29}
\end{equation}
where $N_e^0$ is the number of the incoming electrons.

Therefore, the characteristic scale of the reduced optical
thickness $t$  for the number
of electrons it is determined by the quantity $1/I_{np}$.
This should be compared to the corresponding scale for the
variation of the  electron polarization 
given by  $1/ |\,P_c\,I\,|$.
The relation between these scales is given by the spin asymmetry
\begin{equation}
A(x)= {\sigma_{++} -\sigma_{+-}\over
\sigma_{++} +\sigma_{+-}} = {I\over I_{nl}}\,,
\label{30}
\end{equation}
which is shown in Fig. 2. In the region $x<5$ (which is the most
interesting from the experimental point of view) the spin
asymmetry is small, it does not exceed 10\%
\begin{equation}
|\, A(x) \,| \, < \, 0.09 \;\;\;{\rm at} \;\;\; x\, < \, 5\,.
\label{31}
\end{equation}
In a more wide region $x< 20$ the quantity $A(x)$ does not exceed
30\%.

{\bf 2.} If one is interested only in the mean electron helicity, one
can use a simple alternative approach. Instead of the
variables $N_e$ and $\zeta_z$ one can introduce two new variables
\begin{equation}
N_\pm\,={1\over 2}(1\pm \zeta_z)N_e\, ,
\label{32}
\end{equation}
which are the numbers of electrons with  different signs of
helicity. For these quantities the following equations are valid
\begin{equation}
dN_\pm=-N_\pm\,2n_L\, \sigma_{\pm}dz=-N_\pm\,{\sigma_{\pm} \over
\pi r_e^2} dt\, ,
\label{33}
\end{equation}
where $\sigma_{\pm}$ are the total cross sections for the
scattering of electrons with helicities $\lambda =\pm 1/2$ off the
laser photons (taking into account the laser polarization)
\begin{equation}
\sigma_{\pm}=\pi r_e^2 (I_{np}\pm P_cI) \, .
\label{34}
\end{equation}
Eqs. (\ref{32})--(\ref{34}) are equivalent to Eqs. (\ref{8}),
(\ref{9}) and the last equation of the system (\ref{17}). Let
us stress that the transverse components of the electron polarization
do not appear in Eqs. (\ref{33}).

The described approach could be interpreted as follows:  {\it
(i)} the electron bunch is a mixture of electrons 
with  helicities equal to $+1/2$ and $-1/2$; {\it (ii)} the
numbers of electrons with given helicity change independently; {\it
(iii)} the mean value of the longitudinal polarization is
determined as $\zeta_z\,=\,(N_+-N_-)/(N_+ + N_-)$.  Formula (\ref
{29}) for the number of the undeflected  electrons can be
transformed into a form corresponding to Eq.(\ref{33})
\begin{equation}
N_e= N_{+}^0\,{\rm e}^{-(I_{np}+I)t}\,+N_{-}^0\,{\rm
e}^{-(I_{np}-I)t}\,.
\label{35}
\end{equation}

This approach is satisfactory if one is interested in the
mean value of the electron helicity in the beam only. Besides, the
averaged beam value of the transverse electron polarization is
equal to zero if the bunch has  axial symmetry.  The calculations
in Ref. \cite{Derb} have been performed according to the approach 
outlined in this topic assuming this special assumption.

{\bf 3.} However, this approach is not quite correct since it
only takes into account quantities related to the diagonal
elements of electron density matrix:  $N_\pm=N_e \rho^e_{\pm
\pm}$, but it does not take into account the components of the
electron transverse polarization related to the off-diagonal
elements  $\rho^e_{+-}$ and $\rho^e_{-+}$ of the same matrix.
The inclusion of the transverse electron polarization is necessary,
for example, for the laser conversion of the electron beam into
the high-energy $\gamma$ beam when  multiple Compton
scattering occurs and when the $e \to \gamma$ conversion region
is situated at some distance from the interaction point. In this
case the transverse electron polarization  appears after the
first collision of an electron with a laser photon (even
if this polarization is absent before the collision) and
 affects the angular distribution of the produced
$\gamma$-quanta. The electron polarization (the longitudinal as
well as the transverse one) is varied on the rest path through
the laser bunch. Consequently, an approach which treats the
electron bunch as an independent mixture  of
electrons with  helicities $\lambda=+1/2$ and 
$\lambda=-1/2$, is incorrect for this problem.

The transverse electron polarization was not taken into account,
for example, in Ref. \cite{ZDP}. However, the resulting inaccuracy 
seems to be less than 10 \% as it was noted in (\ref{31}).

{\bf 4.} In conclusion, the variation in polarization for
electrons, which do not scatter, should be
taken into account in simulations of the conversion process. 
It can be realized in the following way:

The electron state is defined by the current values of its energy
$\varepsilon$, the direction of its momentum $z$ and its mean
polarization vector $\mbox{\boldmath $\zeta$}$. The probability
to scatter on the path $dz$ is equal to $dw\, = \, \sigma
(\varepsilon , \zeta_z)\,2\, n_L\, dz$, where
$\sigma(\varepsilon, \zeta_z)$ is the total cross section of the
Compton scattering process. Then, as usual, one can simulate whether the
scattering takes place on this path $dz$ comparing $dw$ with the
random number in the interval (0;1).

If the scattering takes place, one has to simulate the polar
and azimuthal angles using the differential cross section. It
allows to calculate the energy of the scattered $\gamma $-quantum
and its Stokes parameters. A new value of the electron
polarization vector $\mbox{\boldmath $\zeta$}'$ can also be
calculated using the known formulas (see Ref. \cite{KPS}).

If  scattering does not take place, nevertheless, one has to
change the electron polarization vector in accordance with Eqs.
(\ref{17}). It is the essence of the present paper.

\vspace{5mm}

We are grateful to V.~Telnov for discussions concerning problems
of the simulation process which initiate this work. We are also
grateful to V.~Chernyak, I.~Ginzburg, V.~Maisheev, V.~Mikhalev,
A.~Milshtein, A.~Schiller, G.~Shestakov and V.~Strakhovenko for useful
discussions. This work is supported in part by
Volkswagen Stiftung (Az. No. I/72 302) and by Russian Fond of 
Fundamental Research (code 96-02-19114).

\vspace{1cm}

\centerline{{\bf Figure captions}}

\vspace{0.3cm}

{\bf Fig. 1.} The real ($R$) and imaginary ($I$) parts of the Compton
scattering amplitudes at zero angle (see Eq. (\ref{14}) as functions of
 the parameter $x= 4E \hbar \omega_L / (m^2 c^4)$.

\vspace{0.3cm}

{\bf Fig. 2.} The spin asymmetry $A(x)=(\sigma_{++}
-\sigma_{+-}) / (\sigma_{++} +\sigma_{+-})$ for the Compton
scattering function of the parameter $x= 4E \hbar \omega_L /
(m_e^2 c^4)$.

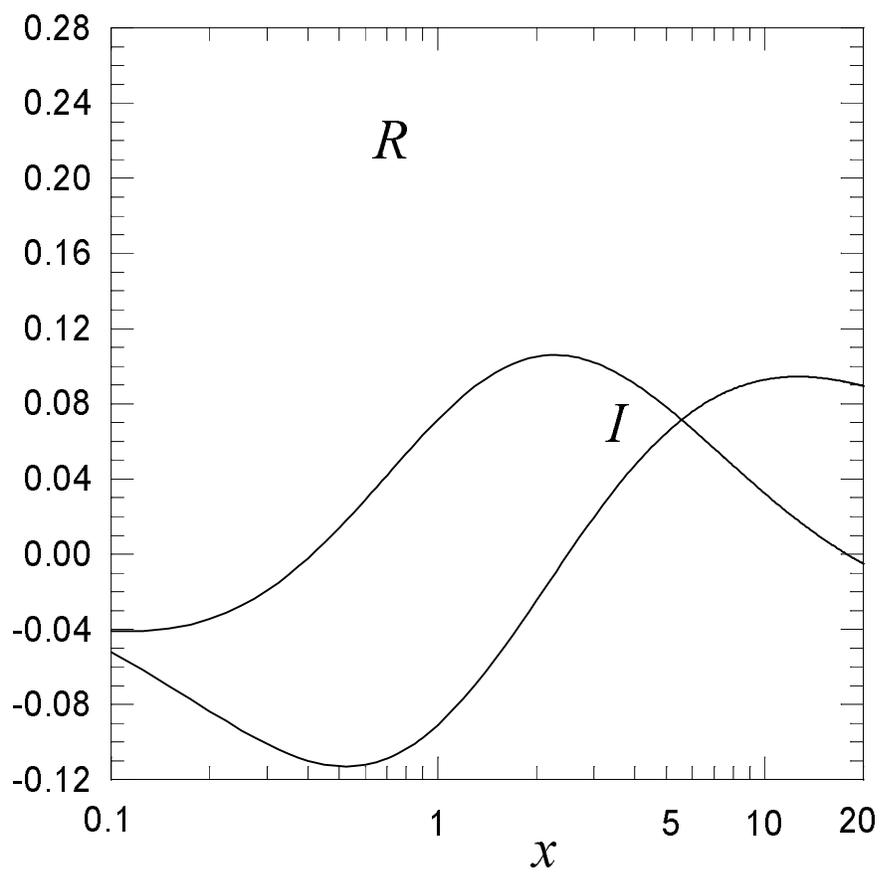
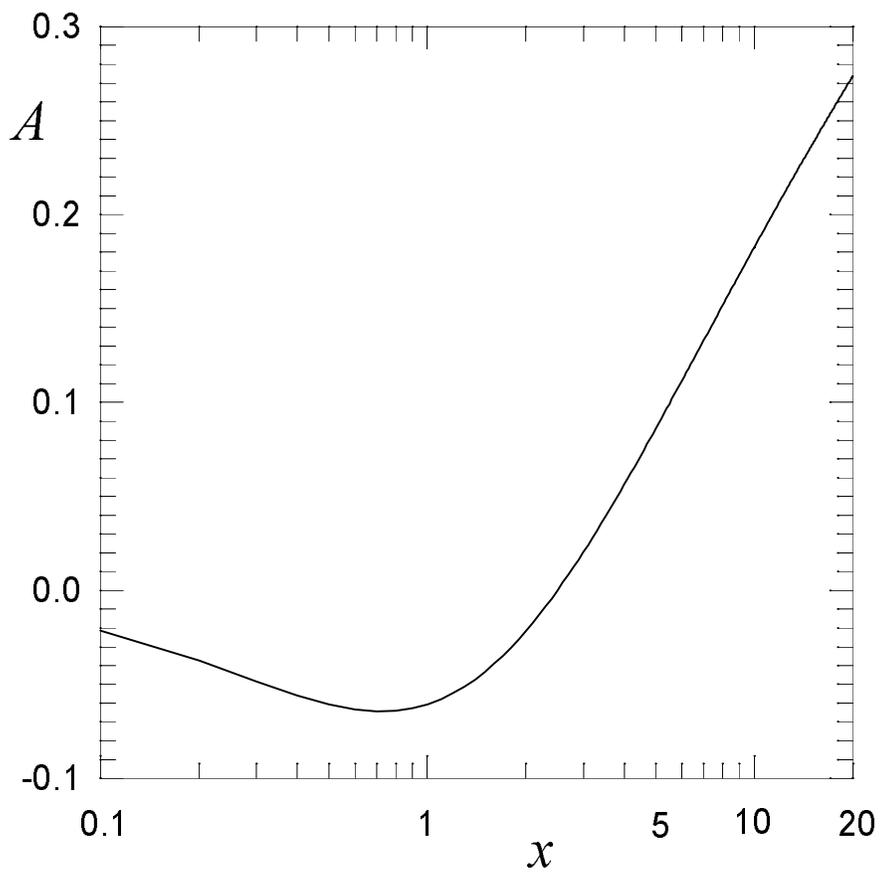
\end{document}